\documentclass[%
preprint,
 amsmath,amssymb,
floatfix,pra,
12pts
]{revtex4-2}

\usepackage[utf8]{inputenc}
\usepackage{graphicx}
\usepackage{dcolumn}
\usepackage{bm}
\usepackage[mathlines]{lineno}
\usepackage{xcolor}
\usepackage{soul}
\usepackage{bm}
\usepackage{bbm}
\usepackage{amsmath,amsfonts,amssymb}
\usepackage{extarrows} 
\usepackage{mathtools}
\usepackage{xr}
\usepackage{hyperref}
\hypersetup{
colorlinks,citecolor=red,
filecolor=blue,
linkcolor=blue,
urlcolor=black
}
\usepackage{xcolor}
\definecolor{sph}{rgb}{0.0588, 0.3216, 0.7294} 
\definecolor{ppk}{rgb}{1.0, 0.4549, 0.0902} 
\newcommand{\nc}{\newcommand}
\nc{\nn}{\nonumber}
\nc{\txt}{\textrm}
\newcommand{\beq}{ \begin{equation} }
\newcommand{\eeq}{ \end{equation} }
\nc{\txtsup}{\textsuperscript}
\nc{\txtsub}{\textsubscript}
\nc{\calL}{\mathcal{L}}
\nc{\U}{\mathcal{U}}
\nc{\T}{\mathcal{T}}
\nc{\E}{\mathcal{E}}
\nc{\cH}{\mathcal{H}}
\nc{\sect}[1]{{\sl #1 .--}}
\newcommand\ket[1]{|#1\rangle}

\nc{\cC}{\mathcal{C}}
\nc{\cI}{\mathcal{I}}
\begin{document}

\title{Quantum transport on networks for supervised classification}

\author{Shmuel Lorber, Oded Zimron, Inbal Lorena Zak, Anat Milo and Yonatan Dubi }
\affiliation{Department of Chemistry, Ben-Gurion University of the Negev, Beer Sheva, 84105,  Israel}


\date{\today}

\begin{abstract}
Classification, the computational process of categorizing an input into pre-existing classes, is now a cornerstone in modern computation in the era of machine learning. Here we propose a new type of quantum classifier, based on quantum transport of particles in a trained quantum network. The classifier is based on sending a quantum particle into a network and measuring the particle's exit point, which serves as a "class" and can be determined by changing the network parameters. Using this scheme, we demonstrate three examples of classification; in the first, wave functions are classified according to their overlap with predetermined (random) groups. In the second, we classify wave-functions according to their level of localization. Both examples use small training sets and achieve over 90\% precision and recall. The third classification scheme is a "real-world problem", concerning classification of catalytic aromatic-aldehyde substrates according to their reactivity. Using experimental data, the quantum classifier reaches an average 86\% classification accuracy. We show that the quantum classifier outperforms its classical counterpart for these examples, thus demonstrating quantum advantage, especially in the regime of "small data". These results pave the way for a novel classification scheme, which can be implemented as an algorithm, and potentially realized experimentally on quantum hardware such as photonic networks.  
\end{abstract}
\maketitle


\newpage
Recent years have seen a huge advance in computational capabilities which are based on machine-learning (ML) algorithms. Classification, pattern recognition, complex predictions and adaptive imitation, among others, are typical problems where ML shows incredible advantage over regular search  or inference algorithms~\cite{alzubi2018machine,shalev2014understanding,bishop2006pattern,gao2017efficient}. However, although ML algorithms work very well and have found a large variety of
applications with unprecedented success, they are limited in a very basic way - they are {\sl emulators}, meaning that they require a digital component (e.g. a transistor) to act as a non-linear device (e.g. to mimic the inherent non-linearity of a neuron). Efforts have been devoted to finding hardware elements which directly show non-linear functionality (for example memristors \cite{thomas2013memristor,gao2021memristor,wang2019reinforcement,xu2021advances, mehonic2020memristors} or other physical implementations \cite{ewaniuk2023imperfect, d2022physics, lee2022mechanical, spall2022hybrid,  wright2022deep}). 

In light of recent advances in quantum computing (and its advantages over classical computation for certain problems \cite{patro2020overview}), various generalizations of ML algorithms to quantum computation have been suggested, \cite{biamonte2017quantum,situ2020quantum,narayanan2000quantum,da2016quantum,li2020quantum,zeng2019learning,farhi2018classification,mitarai2018quantum,dallaire2018quantum,killoran2019continuous,grant2018hierarchical,tacchino2019artificial,wan2017quantum,zhao2019building,he2021variational,beer2020training,steinbrecher2019quantum,cong2019quantum,dalla2020quantum,schuld2014quantum,rebentrost2018quantum,amin2018quantum,tang2019experimental,carleo2017solving,gao2017efficient,cerezo2022challenges, caro2022generalization,cong2019quantum, gao2018quantum, hu2019quantum}  and were applied to various tasks such as quantum encoding, simulating many-body systems,  quantum state discrimination, learning from experiments  \cite{farhi2018classification,wan2017quantum,zhao2019building,he2021variational,beer2020training,steinbrecher2019quantum,cong2019quantum,huang2022quantum}, and supervised classification \cite{Wang2022}. Importantly, these approaches to quantum machine learning consider the "standard" quantum computer architecture, namely qubits and gates. 

Here we take a very different approach, and propose a quantum approach to Supervised classification (SC) , based on quantum transport of particles through a "quantum classification network" (QCN), which in principle can be implemented in current experiments \cite{caruso2016fast}. 

SC is one of the most basic computational tasks performed by machine-learning algorithms \cite{shalev2014understanding}. In SC (in its most simplistic form)  the algorithm is supplied with a training set which  contains a series of tagged inputs (for instance a series of images, each tagged to label the content of the image, e.g. "cat", "dog", etc.). The algorithm uses these images to "train" the system, i.e. to optimize a set of internal parameters. Then, when a test input is supplied, the algorithm can categorize it into its appropriate class.

\begin{figure} 
\centering
\includegraphics[keepaspectratio=true,scale=1.6]{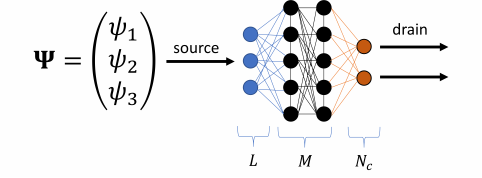}
\caption{\textbf{ Schematic representation of the quantum classification network.} A wave function $\Psi$ of length $L$ ($L=3$ in this example) is encoded into an input excitation of particles into the network, who's parameters are determined following a training protocol. The particles exit the network through one $N_c$ sites, representing the $N_c$ classes. The class of $\Psi$ is then determined by the exit site from which the current is maximal.  }
\label{schematic}
\end{figure}

We provide three fundamental examples for QCN. The first is based on classifying wave-functions according to their overlap with predetermined groups, the second is based on classifying wave functions according to their level of localization. The third example, based on experimentally measured data, is based on classification of substrates according to their reactivity properties. We show that within these examples, the quantum-network based SC outperforms standard machine-learning algorithms under certain conditions, especially for small data sets, thus providing a direct example for quantum advantage. Our proposal can be implemented experimentally on various systems such as networks of quantum dots or quantum wave-guide arrays, thus paving the way for a new design for hardware-based quantum computations.  
\vskip 0.5truecm
\textbf{\large Setup and formulation}\\
\normalsize \textbf{Classification protocol.--} We start by describing the general classification protocol using quantum transport, namely the problem of assigning an input vector $\Psi$ to a specific class $C[\Psi]$. The vectors belong to a Hilbert space of size $L$ (e.g. $L=2$ for qubits), and there are $N$ classes to choose from. Our classification protocol is defined as follows. A network is constructed which has $L$ entry nodes, a "hidden layer" of $M$ nodes and a layer of $N_c$ exit nodes, corresponding to $N_c$ classes (see Fig.~\ref{schematic}). The network is excited such that the excitation is defined by the state $\Psi$ (i.e. a quantum particle enters the network through the source sites in a way which is defined by the state $\Psi$, as described in the following  section). As a result, current flows through the network and exits through the drain sites. From the currents, the class of $\Psi$, $C[\Psi]$ is determined by simply evaluating the currents from the different drain sites, and assigning the class to the node from which current is maximal. 

To give a concrete example (as in Fig.~\ref{schematic}), consider states from an $L=3$ Hilbert space, which can be classified into two classes. A specific state $\Psi$ is encoded into the injection of particles to the network from the source (blue circles in Fig.~\ref{schematic}), and current leaves from the two drain sites (orange circles in Fig.~\ref{schematic}). If most of the current comes out of site $1$ (say the top site), then the state $\Psi$ will be classified as belonging to class $1$ (i.e. $C[\Psi]=1$) and vice versa. This scheme is substantially different from the one proposed in Ref.~\cite{Wang2022}, which involves a local measurement after a stochastic quantum random walk, and requires local control of both Hamiltonian parameters and local dephasing.  

{ \bf The quantum network--} We model the QCN using 
a tight-binding Hamiltonian of the form $\cH=\sum_{i,j} h_{ij} c^\dagger_i c_j+h.c.$, 
where $c^\dagger_i (c_i)$ creates (annihilates) a particle on the network node $i$. 
This is a general description which can describe many quantum transport networks, 
including (but not limited to) electron transport in quantum dots 
\cite{sarkar2022emergence}, exciton transport in bio-networks \cite{zerah2021photosynthetic} and photons in wave-guides \cite{mookherjea2001optical,chen2021tight}. We distinguish between nodes on the entry sites (indexed $i_{in}=1,...,L$), on the network ("hidden") layer (indexed $i=1,...,M$) and exit 
sites (indexed $i_{out}=1,...,N_c$). The network structure is such that entry sites 
are coupled to all network sites but not to each other and not directly to the exit 
sites (and similarly the exit sites are only connected to the network sites). The 
network sites may be interconnected between themselves.

Current propagation through the network, as well as the connection of the network to the environment (i.e. input and output nodes) is modeled using the Gorini-Kossakowski-Sudarshan-Lindblad (GKSL) quantum master equation \cite{lindblad1976generators,gorini1976completely} for the density matrix of the system, $\rho$,  
\begin{eqnarray}
\dot \rho &=&-i[H, \rho] + \Sigma_k( V_k^{\dagger} \rho V_k- \frac{1}{2}(V_k^{ \dagger}V_k \rho + \rho V_k^{ \dagger}V_k))  \\ \nonumber
&=&-i[H, \rho] + \mathcal{L}[ \rho] 
\label{eqn1}
\end{eqnarray}
where $ [ \cdot , \cdot ]$ is the commutator, and $V_k$ are the so called Lindblad operators. In the above setup, k=[1,2] for $V_{in}$ ($V_{out}$) describing the source and drain terms. 

The drain term, $V_{out}$, is defined by a set of $N_c$  $V$- operators describing the extraction of current from the network, of the form $V_{out,n}=\gamma^{1/2} c_{r}$ \cite{zerah2020effects}, where $c_r$ annihilates a particle on the site $r$, where $r=1,...,N_c$ are indices of the exit sites. The source term $V_{in}$ is used to encode the wave-functions that are classified. If a vector $\Psi$ is considered, the encoding is performed by setting $V_{in}=\gamma^{1/2}_{in} \sum^L_{i=1} \Psi(i) c^\dagger_{i}$, where the index $i=1,...,L$ runs over the source sites. 

Once the Hamiltonian is defined (see next section for the training protocol) and the source and drain Lindbladians are provided, we proceed by evaluating the current from each exit node \cite{zerah2020effects} (see Methods section for details)

\textbf{ Network training and validation.--} In order to succeed in classification, the network parameters - in this case, the tight-binding amplitudes in the Hamiltonian - need to be determined according to a given training set. A Training set 
\begin{eqnarray} TS=\left\{ \varphi_{n},C_n\right\},~n=1,...,N_{TS}~~  \nonumber
\end{eqnarray} is a set of input vectors $\varphi_{ n}$ along with their classification $C_n$, where $C_n$ is an integer from 1 to $N_c$, referring to the class of the state $\varphi_{n}$, and $n=1,...,N_{TS}$ is an index running on all the vector states in the training set. 


Once the training set is chosen, the next step is the system training, i.e. finding a Hamiltonian that can achieve the classification described above. According to the classification protocol described above, we define $J_r[\varphi]$ as the current output from the drain site $r=1,..,N_c$ when the input vector is $\varphi$. The {\sl class} of $\varphi$ is thus defined as \begin{equation}\cC[\varphi]=\mathrm{Index}\left( \mathrm{Max}\left( J_r[\varphi],~r=1,...,N_c \right)\right)\end{equation}, i.e. the site index of the drain site with maximal current. 

Training the Hamiltonian thus amounts to minimizing the cost function 
\begin{equation}
CF(\cH,TS)=\sum^{N_{TS}}
_{n=1} \left(\cC[\varphi_n]-C_n \right)^2~.
\label{eq4}
\end{equation}  For the optimization we typically use particle-swarm optimization (PSO) algorithms \cite{poli2007particle}, in combination with standard gradient-descent algorithms, which were found to yield the fastest and most stable convergence.

Once the training is performed, the QCN is validated by applying the algorithm to $N_v$ validation vectors, $\psi_n (n=1,...,N_v)$, which do not belong to the training set but their "true" class, $\cC_n$ can be calculated (or is known externally). Then, by comparing $\cC_n$ to the output class of the algorithm $\cC[\psi_n]$, the performance of the QCN can characterized by the standard measures of classification, namely {\sl precision}, $P$, and {\sl recall} $R$ \cite{shalev2014understanding} (perfect precision $P=1$ means that in the classification process there were no false positives, and perfect recall $R=1$ means there were no false negatives). 

We note on the fly that "training", in the way described above, is not the same as the training defined in modern machine-learning algorithms. Indeed, in our case, the full set of data is apriori required, and "training" implies optimizing the cost function with respect to the given data (rather than updating the system parameters when new data arrives, as in modern machine learning algorithms). In that sense, our protocol is closer to Hebbian learning of a Hopfield network \cite{tolmachev2020new}.  \\

\textbf{\large  Results }\\
\newline\textbf{Example I : group overlap classification.--} 
Our first example entails SC according to the overlap of an input wave-function with different sets of pre-determined groups of known wave-functions. Consider $G$ groups of wave-functions, each group containing $N_G$ wave-functions $\varphi^{g}_{n},~g=1,...,G,~n=1,...,N_G$. The class of an input vector $\psi$ will be determined according to its overlap with the wave-functions of the group $n$. Put in mathematical form, we define the average overlap of $\psi$ with the group $g$, \beq \eta_g=\frac{1}{N_G}\sum^{N_G}_{n=1}| \langle \varphi^g_n | \psi\rangle |^2\eeq (we consider for simplicity only real-valued wave-functions, 
otherwise, in the case of complex-valued wave-functions one can take the absolute value). The classification of $\psi$ is defined as the index of the group of maximal overlap, $\cC[\psi]=\mathrm{Index}[\mathrm{Max}\{ \eta_1,\eta_2,...\eta_{G}\}]$. The quantum network classification is obtained by setting the network to have $G$ exit nodes, and training the network such that when a vector $\varphi^g_n$ is input, most of the current will leave through exit node $g$, i.e. by minimizing the cost function 
\beq CF(\cH,TS)=\sum^G_g \sum^{N_G}_n \sum^{G}_i (J_i[\varphi^g_n]-\delta_{i,g})^2 \eeq, where $i$ is the index of the exit node and $\delta$ is the Kronecker delta function.  

We start our demonstration of QCN by examining the simplest possible  case of $L=2, G=2, N_G=2$, namely by classifying inputs (of length $2$, i.e. qubits) according to their overlap with only two vectors. The training set of the two vectors, $\varphi_{1,2}$, is given by
\begin{eqnarray}
TS=
\begin{cases} \label{eq1}
\varphi_{1} = \cos(x) \ket{0} + \sin(x) \ket{1} , C_1 = 1  \\
\varphi_{2} = \sin(x) \ket{0} + \cos(x) \ket{1} , C_2 =2   
\end{cases}
\end{eqnarray} where we chose $x=0.3$ (other values give similar results). A network of 2-4-2 was chosen (namely, 4 nodes in the "hidden layer", $M=4$), and trained (by varying the hopping matrix elements such that the cost function is minimized) using PSO as mentioned above.

Once the training is complete, we tested the classification using validation set containing 1000 random states $\psi_n$, each assigned a class according to its overlap in the following way. Defining the overlap $\eta^{(n)}_g=\langle \psi_n| \varphi_g \rangle $, the class of $\psi_n$ is $\cC_n=1$ (or 2)  if $\eta^{(n)}_1 >\eta^{(n)}_2$ (or $\eta^{(n)}_1 <\eta^{(n)}_2 $). The class $\cC[\psi_n]$ defined by the QCN is the index of the exit site which carries the most current when $\varphi_n$ is inserted into the network. 

\begin{figure}
    \includegraphics[width=0.48\linewidth]{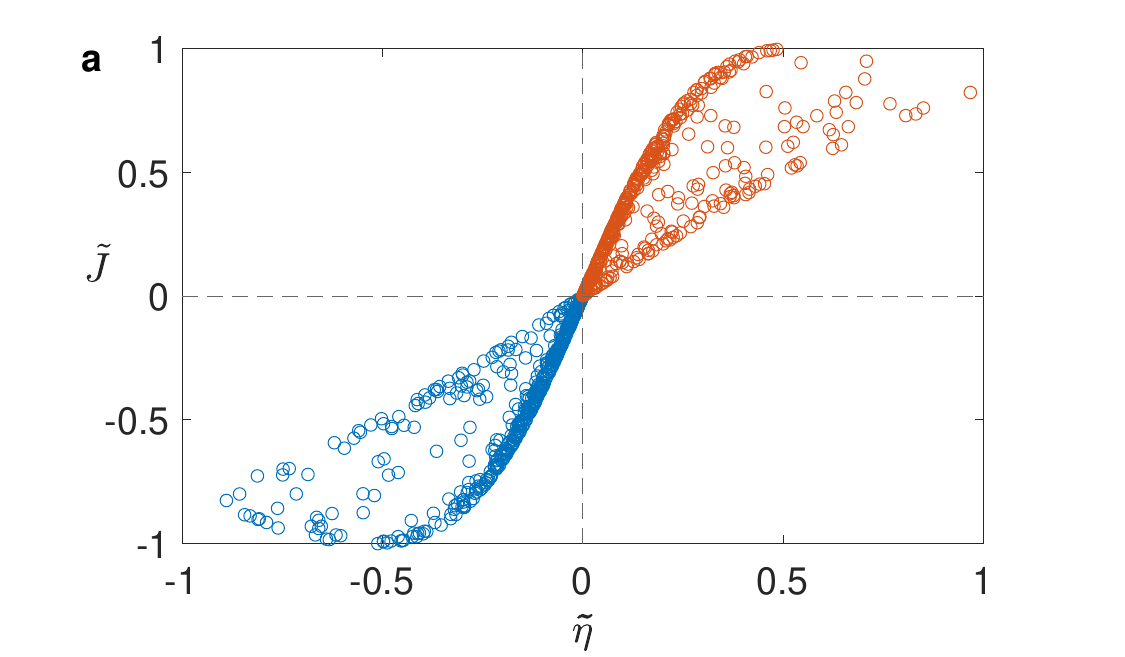}
   \vspace{0.5cm} 
    \includegraphics[width=0.48\linewidth]{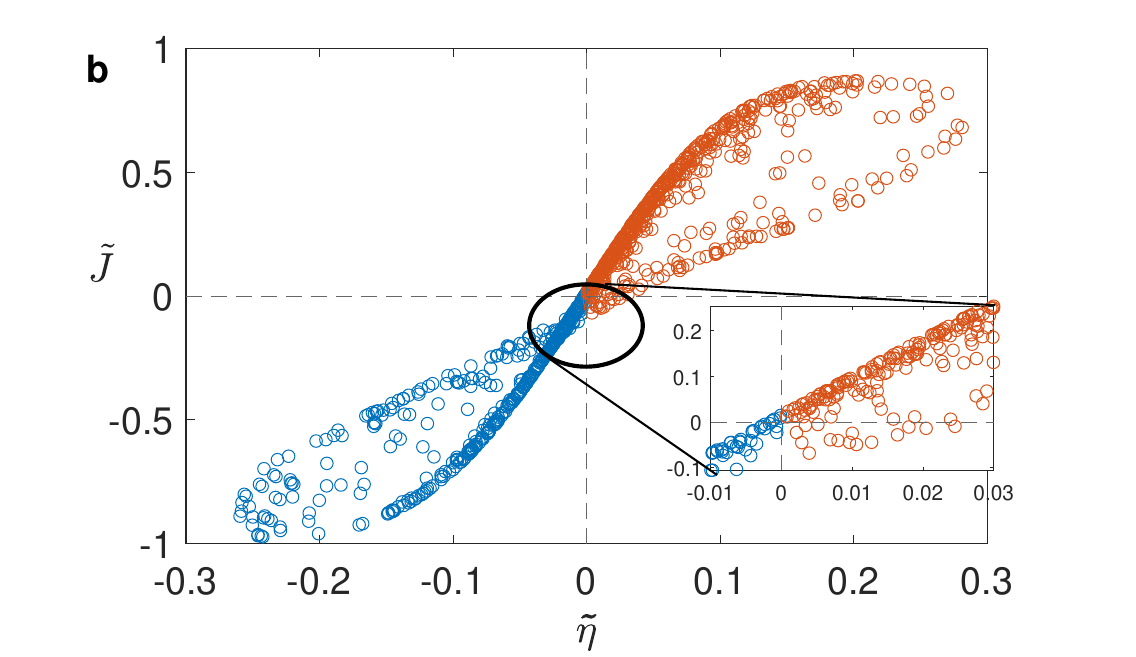}
\caption{ {\bf Performance of the QCN protocol for group overlap classification. } {\bf a}: Current balance $\tilde{J}$ vs overlap balance $\tilde{\eta}$ for $N_v=1000$ validation vectors for $G=2, ~N_G=2$. No points reside on the 2nd and 4th quarters, indicating that $\tilde{J}$ and $\tilde{\eta}$ always have the same sign. This means perfect classification for this case, with $P=R=1$. {\bf b}: Same as {\bf a}, for $G=2,~N_G=40$ (a training set consisting of 2 groups with 20 vectors each). We find $P=0.9914$ and $R=0.9745$, almost perfect classification. The outliers are due to validation vectors which have very close overlap to the two classes, and thus lie close to the $\tilde{\eta}=0$ point (see enlargement of this area in the inset).}
\label{fig2_current_overlap}
\end{figure}

For this example, we find $P_2=R_2=1$, i.e. {\sl perfect} classification. Put simply, there are no validation vectors $\psi_n$ which belonged to class 1 and, when injected to the trained network, had most current come out of node 2 (and vice versa). This is depicted in Fig.~\ref{fig2_current_overlap}a. For simplicity, we define for each $\psi_n$ an overlap balance $\tilde{\eta}_n=\frac{\eta^{(n)}_1-\eta^{(n)}_2}{\eta^{(n)}_1+\eta^{(n)}_2}$ and a current balance $\tilde{J}_n=\frac{J_1[\psi_n]-J_2[\psi_n]}{J_1[\psi_n]+J_2[\psi_n]}$. Both $\tilde{\eta}_n$ and $\tilde{J}_n$ vary between $\pm \frac{1}{2}$, and perfect classification means that they have the same sign for every $\psi_n$. In Fig.~\ref{fig2_current_overlap}a we plot $\tilde{J}_n$ vs $\tilde{\eta}_n$ for all the validation vectors. Indeed, for all $\psi_n$'s, the sign of $\tilde{J}$ and $\tilde{\eta}$ is the same. 




Of course, this is an extremely simple example. To generalize it, we have extended the study to the cases of $N_G=20$ and $N_G=40$ (still with $G=2$, i.e. two classes each containing $N_G$ vectors). The states were of the same form, $\varphi_i=\cos(x_i)|0\rangle+\sin(x_i)|1\rangle$, where $x_i$ were chosen randomly from a uniform distribution. Using the same training protocol as for the $N_G=2$, we find $P_{20}=0.9961,~R_{20}=0.9644$ and $P_{40}=0.96767,~R_{40}=0.9984$, i.e. almost perfect classification. The current balance $\tilde{J}$ vs overlap balance $\tilde{\eta}$ for $N_G=20$ is shown in Fig.~\ref{fig2_current_overlap}b. The outliers, namely validation states which give a false positive or negative, are such that have an extremely similar overlap between the two training sets, and thus lie close to the $\tilde{\eta}=0$ point, as shown in the inset to Fig.~\ref{fig2_current_overlap}b. The specific shape of the plot arises due to the specific choice of vectors in the overlap classes, Eq.~\ref{eq1}. Similar tests for other random choices of vectors give similar results. 

\begin{figure}
 \includegraphics[width=1\linewidth]{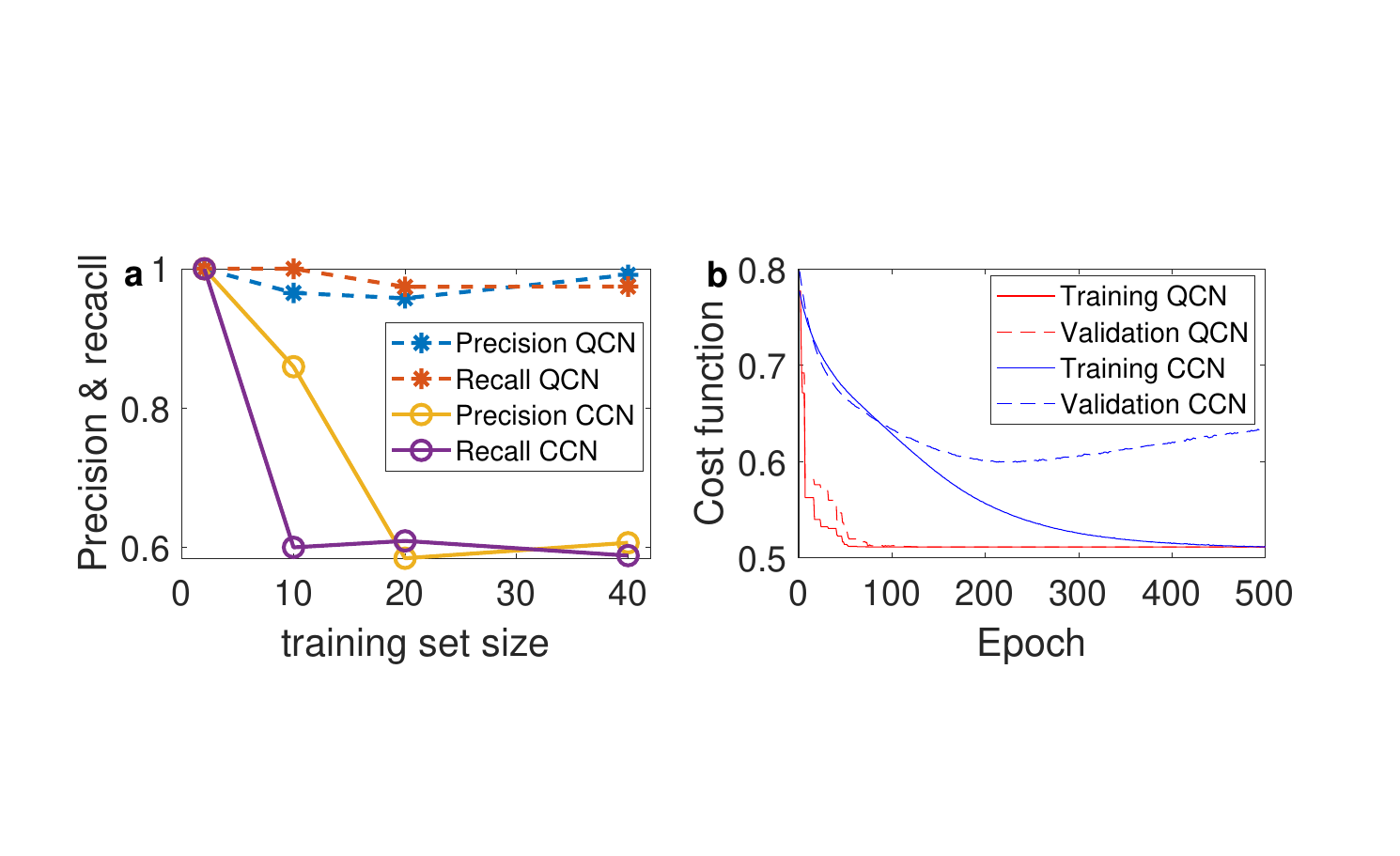}
   \vspace{-2.5cm} 
  \caption{{\bf Quantum advantage demonstrated on overlap-based classification.} (a) Precision and recall with varying sizes of training sets of the QCN (stars) and CCN (circles). As can be seen, except for $N_{TS}=2$, the QCN algorithm shows advantage over the classical algorithm. (b) Cost function vs training epoch of the training and validation sets (solid and dashed lines, respectively) for the QCN and CCN (red and blue, respectively), for an $N_G=40$ overlap classification. While for the quantum network the validation set cost function follows the training set cost function, the classical validation cost function diverges, a clear indication of "overfitting" due to the small training set.}
  \label{fig:OverlapCombined}
\end{figure}

To see how the QCN compares to a classical classification network (CCN), we ran the same classification problem with a standard classical neural network (see Methods section for details). While for $N_G=2$ the classical algorithm performed perfectly, it provided rather poor classification for $N_G=20, 40$. In Fig.~\ref{fig:OverlapCombined} we plot $P$ and $R$ for $N_G=2,20,40$ for the QCN (stars) and classical algorithm (circles). Solid and dashed lines are guides to the eye. It is clearly visible that the QCN substantially outperforms the CCN in this regime. However, as we go to much larger training sets (e.g., $N_G=600$, not shown) the classical algorithm gives $P=0.9853, R=1$. This affirms that the quantum advantage we observe is limited to small training sets.  


To further elaborate on the quantum advantage of the QCN, in Fig.~\ref{fig:OverlapCombined}b the cost function is plotted vs the number of iterations (so-called calculation epoch) of the training and validation sets for the QCN (red solid and dashed lines, respectively) and the CCN (blue solid and dashed lines, respectively), for $N_G=40$. For the QCN the validation set follows the training set behavior, which is what is expected from a well-performing classification network. For the classical network, on the other hand, the validation set begins to diverge, leading to poor classification. This behavior is a hallmark for the so-called "overfitting" problem \cite{dietterich1995overfitting, ying2019overview}. It thus seems that the QCN overcomes the overfitting problem for small data sets. We raise here a conjecture that for this problem, finding the minimum of the cost function on a (quantum) Hilbert space is a convex optimization problem, but becomes non-convex for the classical problem, which leads to the overfitting. Proof of this conjecture is left to future studies. 


As a final example for overlap classification, we performed the calculation for $G=4$, namely 4 classification classes. We use a 3-5-4 network. The results for the precision and recall (averaged over the 4 classes) is given in table 1 below. Comparing to the classical algorithm, we again find substantial quantum advantage for the QCN. 


\begin{table}[h]
\begin{tabular}{ |c||c|c|c|  }
 \hline
& $N_{G}=4$ & $N_{G }=20$& $N_{G}=40$\\
 \hline
 Precision QCN   & 0.9309 & 0.9116& 0.9383\\
 Recall QCN&   0.9000 & 0.8888& 0.8646\\
 Precision CCN   & 0.5002 & 0.6537& 0.9100\\
 Recall CCN&   0.2874 & 0.6848 & 0.6533\\
 \hline
\end{tabular}
 \label{table.1}
\caption{Classification success values for $G=4$ }
\end{table}

We point out that this example, although simple to understand, is a somewhat atypical classification problem. The reason is that in this example, the classes and the training set are defined by the same group of input vectors. We thus proceed with a more standard example in the next section.  

\textbf{Example II : classification by level of localization.--} The next classification assignment the QCN was trained to achieve is classification based on level of localization. We measure the level of localization by the Inverse Participation Ratio (IPR) \cite{hikami1986localization,fyodorov1993level,evers2000fluctuations}, which is a well-known measure of localization defined by 
\begin{eqnarray}
\cI[\psi]=\left(\sum^{L}_{i=1}|\psi(i)|^4\right)^{-1}~~.
\end{eqnarray}

The IPR ranges from $\cI=1$ for a completely localized wave-function to $\cI=L$ for a fully delocalized wave-function (where the weight is equal over all basis state). Using different measures of localization (e.g. weave-function entropy) yields similar results.   

As a first example, we set $L=5$ and randomly chose $N_{TS}=10$ states, divided into $G=2$ groups of "localized states" with $\cI<2$ and "extended states" with $\cI>3$ (states with $2<\cI<3$ were discarded). After training a 5-8-2 network, 1000 validation wave-functions were tested. 

\begin{figure}

    \includegraphics[width=\linewidth]{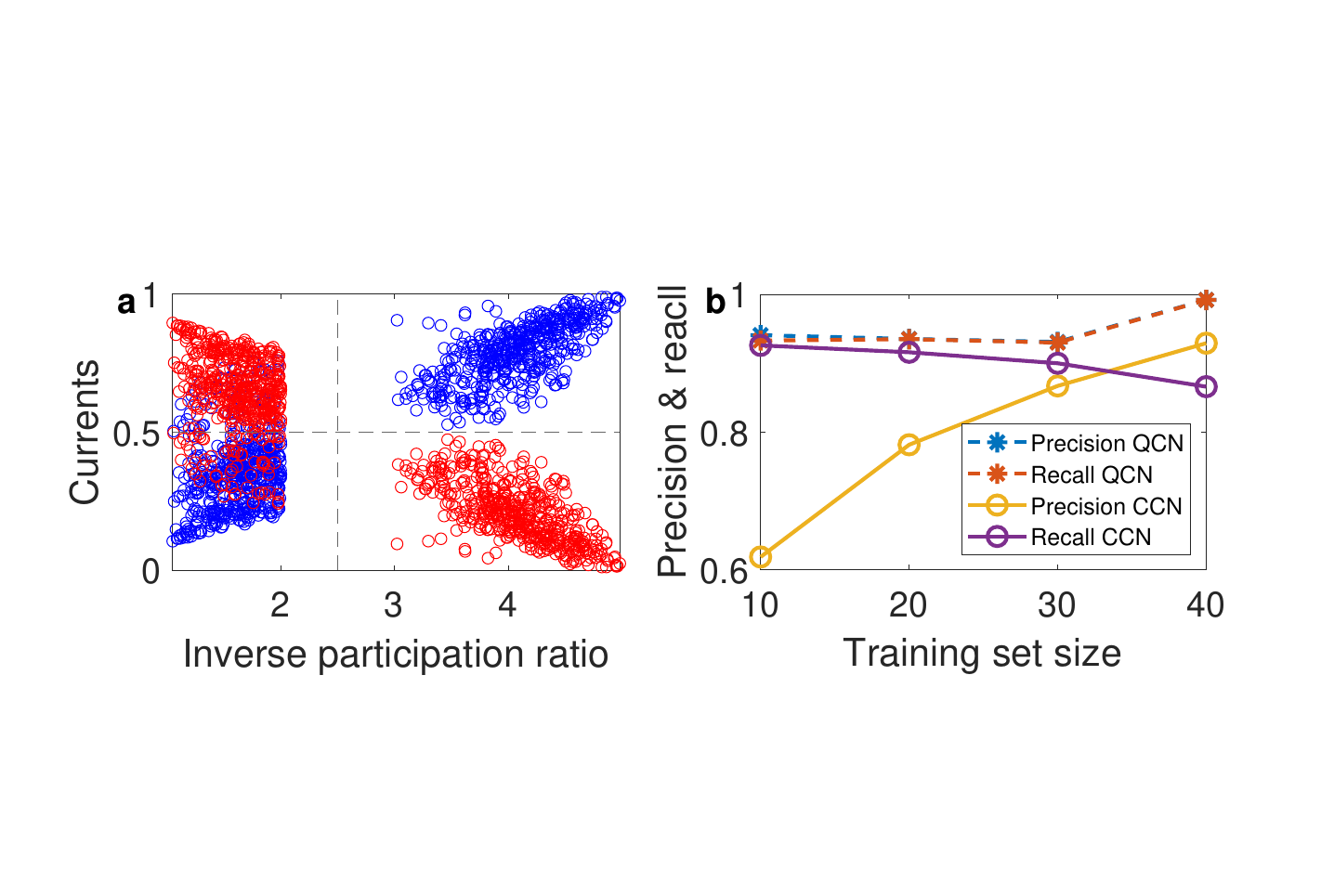}
  \vspace{-2.5truecm}\caption{{\bf  Performance and quantum advantage of localization-based classification and quantum advantage}. (a) Currents from exit node 1 (red circles) and 2 (blue circles) for 1000 wave-functions, plotted against the wave-functions IPR, $\cI$. A clear ability to classify between localized ($\cI<2$) and delocalized $\cI>3$ wave functions is seen - for localized states most of the currents exit from node 1, and for delocalized states most of the current exists from site 2. (b) Comparison of precision and recall of quantum and classical network classification based on wave-function level of localization (via the IPR, see text)  with varying sizes of $N_{TS}$. Data clearly shows the advantage of the QCN }
  \label{fig:IPRcombined}
\end{figure}

In Fig.~\ref{fig:IPRcombined} we plot the current from exit node 1 (red circles) and exit node 2 (blue circles) as a function of the IPR of the test wave-function. It can be clearly seen that for validation states with $\cI<3$ most of the current flows out of node 2 and vice versa, as indeed required for this classification. Precision and recall for this case study were found to be $P=0.9409, ~R=0.9330$. 

As for the case of overlap classification, we proceed to compare the performance of the QCN vs the classical algorithm for different number of training set vectors $N_{TS}$. As seen in Fig.~\ref{fig:IPRcombined}(b), where the precision and recall are plotted as a function of $N_{TS}$ for the QCN (stars) and classical algorithm (circles), the QCN consistently outperforms the classical algorithm over the entire range of examined $N_{TS}$.



\begin{figure} 
\includegraphics[keepaspectratio=true,scale=0.5]{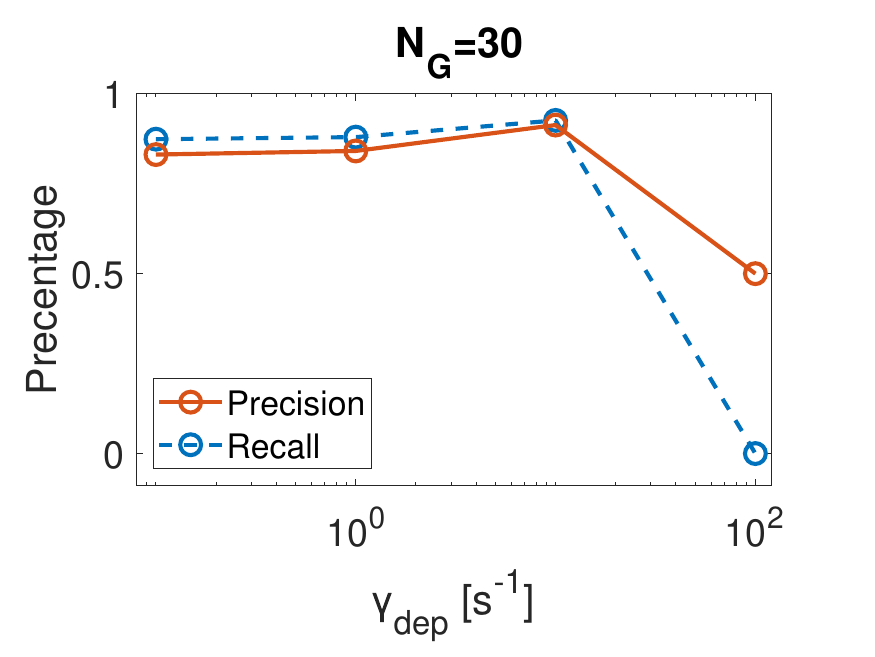}
\caption{{\bf Robustness against dephasing}. Precision and recall of group localization classification for $N_{TS}=30$ as a function of dephasing rate.   }
\label{fig:dephasing}
\end{figure}

\textbf{Robustness against dephasing.--} 
Dephasing, or the general loss of wave-function coherence, may destroy the ability of the network to perform classification, since the classification depends on the coherent transport of the excitations through the trained network. To explore the effect of dephasing, one can ad a Zeno-type local dephasing terms to the Lindbladian \cite{Breuer2002} with the $V-$operators $V_i=(\Gamma_{\mathrm{dep}})^{1/2} c^\dagger_i c_i$, where $i$ runs over all network sites and $\Gamma_{\mathrm{dep}}$ is the dephasing rate. In Fig.~\ref{fig:dephasing} we plot the precision and recall for a group IPR classification for $N_{TS}=30$ (same calculation as in Fig.~\ref{fig2_current_overlap}) as a function of dephasing rate $\Gamma$ (where $\Gamma$ is taken in units of the average hopping matrix element of the Hamiltonian $\overline{t}$). We find that while the precision actually increases with dephasing rate, the recall decreases substantially as $\Gamma\sim 100 \overline{t}$. One can evaluate this rate for realistic systems. Noting that $\overline{t}\sim 1~\mu$eV for quantum dots \cite{sarkar2022emergence} and $\overline{t}\sim 1 $ eV for photonic mazes \cite{chen2021tight}, it follows that the dephasing time should be longer than $\sim50$ ns for quantum dots and $\sim50$ fs for photonic mazes, which is well within reach in current experimental setups. Similar calculations with different parameters (e.g. different $G$ and $N_{TS}$) gave similar results. \newline

\textbf{\large Implementation of QCN on chemical data}\\
\newline
The two previous examples, while representative, are somewhat artificial - they were created to test the QCN algorithm. Next, we demonstrate the action of the QCN on real experimental data. For this, we turn to the world of chemistry and catalysis, where the use in machine-learning algorithms is beginning to emerge, yet many problems are characterized by having a small data set, i.e. "Small Data" problems \cite{butler2018machine,shalit2023small}.  

Specifically, we consider the classification of aromatic aldehyde substrates, based on their preferred reaction conditions to a deuteration reaction. The aldehydes are classified on a scaling reactivity, into 3 broad categories: (1) those that undergo full deuteration with a catalyst named SIPr, but due to their reactivity tend to proceed and form benzoin. Suppressing this further reaction demands adding also boronic acid (BA) to the reaction. (2) Aromatic aldehydes that undergo full deuteration with SIPr alone, and (3) those that require a triazolium catalyst (TAC) for their full deuteration. The first group contains the most reactive substrates, while the 3rd group contains the least reactive ones. The substrate reactivity depends on various properties, including steric hindrance, dipole charges, partial charges obtained from natural population analysis (NPA) \cite{reed1985natural}, etc. Since no formal method for determining of specific reactivity conditions is known, using machine learning algorithms may prove useful. 



In the experiment, 60 substrates (i.e. molecules) were tested, and experimentally classified into one of the three classes described above. We use these data for the classification scheme as follows. Each substrate is characterized by a list of $10$ numerical values corresponding to its physical characteristics (some stated above, see SI for the full data), which display a broad distribution over the substrates. 
The values are then normalized such that they are distributed in the range $[-1,1]$, and then constructed into a "normalized wave function", namely a list of normalized parameter values, normalized by itself to unity. With $3$ classes and $10$ numerical characteristics, the network thus has $10$ input nodes (representing the numerical characteristics of the substrate) and $G=3$ output nodes (representing their reactivity).

We note that since we are working with  a very small data set, the typical procedure for classification protocols, namely dividing the data set into a training set and validation set, and determining the performance of the protocol from the chosen training set, may be inadequate. We therefore proceed with a slightly more detailed analysis. We focus on accuracy, $A$, defined as the number of correct predictions divided by the total number of predictions \cite{shalev2014understanding}, rather than precision and recall, because of the small data set size. For an estimation of the protocol accuracy, we thus repeat this procedure $10$ times, randomly choosing a validation set and a training set at each repetition, and evaluate the accuracy of the protocol for each realization. This gives us an average accuracy and a standard deviation, which is a better statistical description for the protocol performance. 

\begin{figure}
    \includegraphics[width=0.48\linewidth]{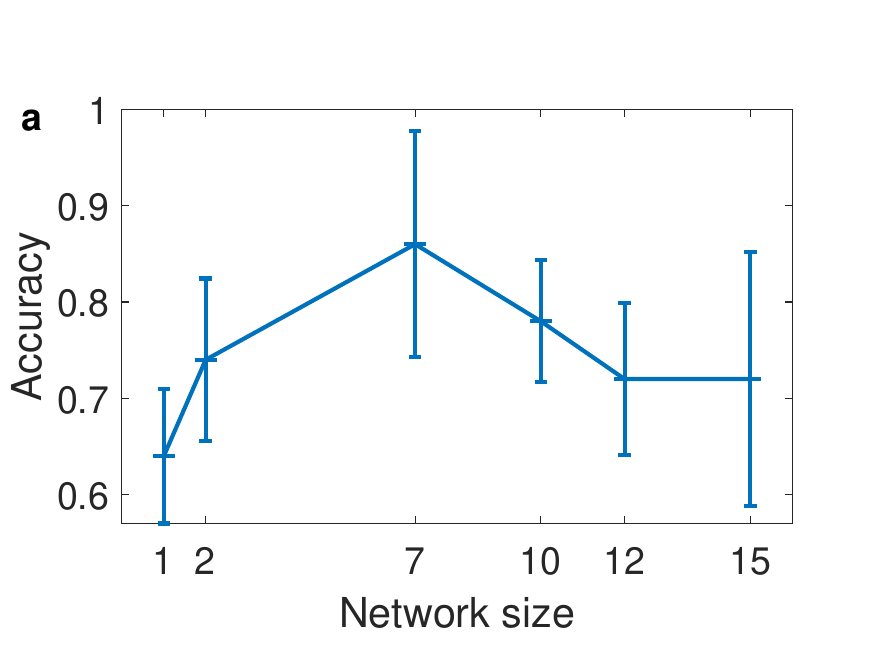}
   \vspace{0.5cm} 
    \includegraphics[width=0.48\linewidth]{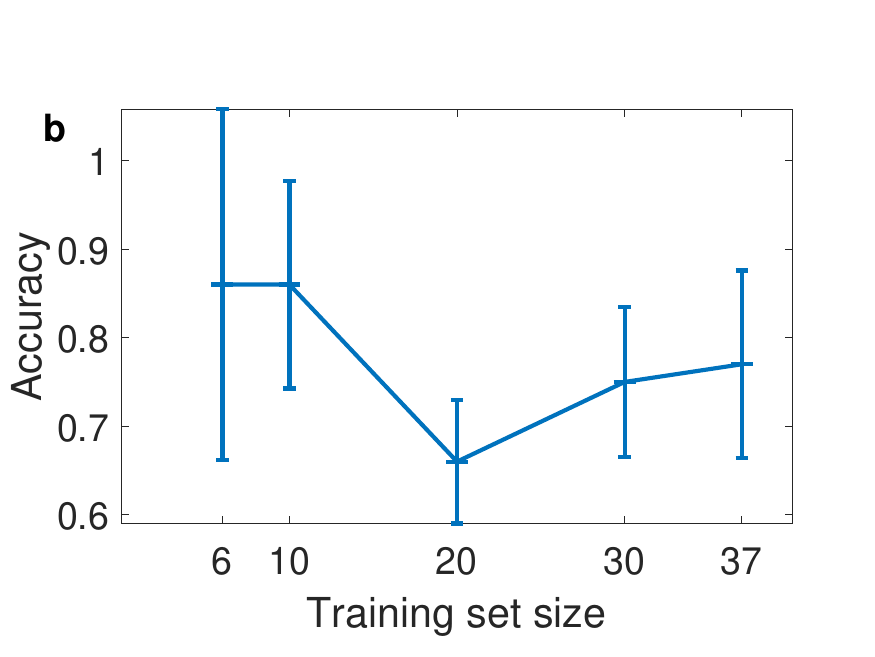}
        \caption{\textbf{Performance of the QCN protocol on real-world data.} The classification protocol (see text for details) is used for randomly chosen training and validation sets, from which the average and standard deviation of the accuracy are evaluated. (a) Average classification accuracy for a network of constant $N_{TS}=10$, as a function of the size $M$ of the intermediate (hidden) layer (see Fig.~\ref{schematic}). The network that presents the best performances is a $10-7-3$ network. (b) Mean accuracy of the QCN classification for different numbers of training set vectors, $N_{TS}$. Best result was obtained for $N_{TS}=10$, with accuracy $A=0.86$, and reaching up to $A=1.0$ for some validation sets. The fact that the best $N_{TS}$ and $M$ are so small shows the potential of the QCN protocol for small data problems. Error bars mark the standard deviation}
        \label{NetworkDimensions}
\end{figure}

The first step is to determine the optimal network size. We select a training set of size $N_{TS}=10$ (and a validation set with $10$ vectors), and repeat the calculation 10 times. In Fig.~\ref{NetworkDimensions}a we plot the accuracy (mean and standard deviation) as a function of the size of the internal layer (e.g. black circles in Fig.~1). We find that a network with 7 sites gives the best accuracy, $\sim 86\%$. 
We then turn to determine how many vectors the training set requires to obtain good classification accuracy. Using the optimal network size found above, Fig.~\ref{NetworkDimensions}b shows the accuracy as a function of training set size (for each size, 10 different realizations were chosen and the accuracy mean and standard deviation were evaluated). Surprisingly, we find that the best accuracy was actually reached for a rather small training set size $N_{TS}=10$, and that increasing the training set size actually reduced the accuracy on average. Similar results were obtained for the precision and recall (see SI). 

The results of Fig.~\ref{NetworkDimensions} show that the QCN protocol averages at around an accuracy of $A=0.86$, and for some validation sets reaches up very close to $100\%$ accuracy (similar values were obtained for the precision and recall). These values are comparable to (and even higher from) the accuracy of similar calculations performed with classical algorithms \cite{butler2018machine,segler2017neural,wu2018moleculenet, liu2017retrosynthetic}, yet were obtained with a much smaller training set (of only $\sim10$ vectors). This demonstrates the possibility of quantum advantage in the QCN protocol in the regime of small data.

\vskip 0.5truecm \textbf{Discussion}\\

Admittedly, the QCN algorithm we present is computationally more expensive, and for large system sizes may be more time consuming, than the classical algorithms. However, it show a "quantum advantage",  especially relevant for the regime of small data \cite{cui2021machine}, where available datasets are limited in size, a regime which seems highly relevant in fields such as material design and chemistry \cite{liu2020small, haraguchi2021size,drechsler2020combining,butler2018machine}. This may become practical for situations where calculation time is an unimportant factor but recall and precision are important,

What is the origin of the quantum advantage of the QCN algorithm? one possible conjecture is as follows. The cost-function of the QCN algorithm is based on the relation between the currents and the hopping matrix elements in a network, a relation which is typically of Lorentzian nature. This means that the optimization landscape is fairly shallow, i.e. the local minima are not "exponentially deep" but "Lorentzianly deep", which allows the algorithm to effectively escape local minima, and reach the global minimum even for a small training set. How come this works for such small training sets? these questions are left for future investigations. 

Importantly, the QCN scheme we present here is not only an algorithm to be implemented on a computer. Rather, it is a scheme that can in principle be implemented using current state-of-the-art experiments using photonic networks (e.g., \cite{viciani2016disorder,caruso2016fast}). Thus, this work paves the way for a quantum machine aimed at performing specific machine-learning computational tasks, beyond the standard paradigm of quantum computing.

\vskip 0.5truecm

 \textbf{\large Methods}\\

\textbf{Current from the Lindblad equation}

As discussed in the Results section, the system is described by a general tight-binding Hamiltonian  and the Lindblad equation,
\begin{eqnarray}
\dot \rho &=&-i[H, \rho] + \Sigma_k( V_k^{\dagger} \rho V_k- \frac{1}{2}(V_k^{ \dagger}V_k \rho + \rho V_k^{ \dagger}V_k))  \nonumber\\ 
&=&-i[H, \rho] + \mathcal{L}[ \rho]~. \nonumber
\end{eqnarray}
We limit the calculation to a single-exciton manifold (we expect only minor changes if one uses the full Fock space, as demonstrated in \cite{zerah2020effects}.

We solve the Lindblad equation and evaluate the current at the steady state. i.e. $\dot \rho_s =0$ \cite{guo2016molecular,zerah2021photosynthetic,zerah2015enhanced,zerah2019environment}

Once $\rho_s$ is at hand, we evaluate the current, $J$, from the extraction sites $J_{i}=\frac{d\langle n_i\rangle}{dt}$ where $\langle n_i\rangle$ is the on-site density ($i$ is the site identifier) , via the relation $\langle n_i\rangle = Tr(\hat{n_i}\rho)$. Substituting the expression for the on-site density into $J_i$ gives an equation for the current,
\begin{eqnarray}
    J_{ij}=\frac{d\langle n_i\rangle}{dt} = \frac{d}{dt}Tr(\hat{n_i}\rho)
\end{eqnarray}
Inserting the derivative into the trace and applying it on its contents yields
\begin{eqnarray}
    J_{i}=Tr(\dot{\hat{n}}_i\rho_s + \hat{n}_i \dot\rho_s)
    \label{eq2}
\end{eqnarray}

Considering only exit sites, this expression formally becomes
\begin{eqnarray}
     J_{ext} = Tr(\hat{n}_{ext}(-i[\cH,\rho_s]+\mathcal{L}_{ext}[\rho_s])) = 0
    \label{eq3}
\end{eqnarray}

The expression above vanishes due to trivial current conservation at the steady state (the two terms are identical in magnitude and opposite in sign). In order to evaluate the exiting current, we thus evaluate only the term relating to the exit sites (or more specifically, to the measuring apparatus), e.g., the term 

\begin{eqnarray}
    J_{ext} = Tr(\hat{n}_{ext} \mathcal{L}_{ext}[\rho_s])~~.
\end{eqnarray}\\

\textbf{Training the classical classification network}
In order to compare the QCN and the classical one, we use the following procedure. First, we take exactly the same training set for both cases. Then, we train the classical network using "Keras" (https://keras.io/), which is a convenient application programming interface (API) for implementing machine learning models in Python. Activation function of the CCN
was "Relu", output layer activation was a "sigmoid", the optimizer
of the network was "Adam", and loss function used was "crossentropy".

\vskip 0.5truecm
\textbf{\large Data availability}\\
No datasets were generated or analysed during the current study. Please contact the corresponding author(s) with questions or
concerns.
\vskip 0.5truecm
\textbf{\large Code availability}\\
The custom code that was created during the work that led to the main results of this article is published in a public GitHub repository: https://github.com/ShmueLorber/Quantum-Classification

\vskip 0.5truecm
\textbf{\large Author contributions}\\
Y.D. conceived the project and supervised the calculations. S.L. designed and performed the calculations. O.Z. performed some calculations and assist in data analysis. I.L.Z. and A.M. supplied the chemical data and performed the chemical experiments to generate the data. All authors participated in the analysis of the data. S.L. and Y.D. wrote the manuscript, and all authors commented on the manuscript. 

\vskip 0.5truecm
\large \textbf{Competing interests}\\
The authors declare no competing interests.


%

\end{document}